\documentclass[final,1p,times]{elsarticle}

\usepackage{graphicx}
\usepackage{amssymb}
\usepackage{amsmath}
\usepackage{amsfonts}
\usepackage{bbm}
\usepackage{braket}
\usepackage{hyperref}
\usepackage{lineno}
\usepackage{xcolor}

\biboptions{sort&compress}

\journal{Annals of Physics Special Issue: Localisation 2020}

\begin{document}

\begin{frontmatter}


\title{Quench dynamics of quasi-periodic systems exhibiting Rabi oscillations of two-level integrals of motion}

\author[1,2]{Leonardo Benini}
\author[3,4]{Piero Naldesi}
\author[1]{Rudolf A.\ R\"{o}mer}
\ead{r.roemer@warwick.ac.uk}
\author[2]{Tommaso Roscilde}

\address[1]{Department of Physics, University of Warwick, Coventry, CV4 7AL, UK}
\address[2]{Univ de Lyon, ENS de Lyon, Univ Claude Bernard, and CNRS, Laboratoire de Physique, F-69342 Lyon, France}
\address[3]{Universit\'e  Grenoble-Alpes,  LPMMC and  CNRS, F-38000  Grenoble,  France}
\address[4]{Institute for Quantum Optics and Quantum Information of the Austrian Academy of Sciences, Innsbruck, Austria}

\begin{abstract}
The elusive nature of localized integrals of motion (or \emph{l-bits}) in disordered quantum systems lies at the core of some of their most prominent features, i.e.\ emergent integrability and lack of thermalization. Here, we study the quench dynamics of a one-dimensional model of spinless interacting fermions in a quasi-periodic potential with a localization-delocalization transition. Starting from an unentangled initial state, we show that in the strong disorder regime an important subset of the $l$-bits can be explicitly identified with strongly localized two-level systems, associated with particles confined on two lattice sites. The existence of such subsystems forming an ensemble of nearly free $l$-bits is found to dominate the short-time dynamics of experimentally relevant quantities, such as the Loschmidt echo and the particle imbalance. We investigate the importance of the choice of the initial state by developing a second quench protocol, starting from the ground-state of the model at different initial disorder strengths and monitoring the quench dynamics close to the delicate ETH-MBL transition regime. 
\end{abstract}

\begin{keyword}
Localization \sep Non-equilibrium dynamics \sep Many-body systems

\end{keyword}

\end{frontmatter}

%%%%%%%%%%%%%%%%%%%%%%%%%%%%%%%%%%%%%%%%%%%%%%%%%%%%%%%%%%%%%%%%%%%%%%%%%%%%%%%%%
\section{Introduction}
%%%%%%%%%%%%%%%%%%%%%%%%%%%%%%%%%%%%%%%%%%%%%%%%%%%%%%%%%%%%%%%%%%%%%%%%%%%%%%%%%
The phenomenon of quantum localization~\cite{Akkermans-book, Anderson-book, Anderson1958AbsenceLattices} is one of the most fundamental contributions that quantum mechanics offers to the description of many-body dynamical behavior. It has wide-ranging implications in many foundational fields of physics, and in particular in the context of quantum statistical mechanics. The extension of Anderson localization (AL) to the interacting regime, i.e.\ many-body localization (MBL) \cite{NandkishoreH2015,Alet2018,Abaninetal2019, GOPALAKRISHNAN2020}, implies the breakdown of the ergodic hypothesis for a large family of generic (not fine-tuned) models of quantum matter. It was pointed out early on \cite{Serbynetal2013}, and then investigated further \cite{Huseetal2014, Imbrie2016, Imbrieetal2017}, that the fundamental mechanism governing the failure of thermalization in such systems is the emergence of a \emph{local} form of integrability, namely the appearance of an extensive number of locally conserved quantities at strong disorders. In the case of non-interacting systems, the emergent integrability is simply related to the conservation of particle numbers in the localized eigenstates of the model Hamiltonian.  In MBL systems, the dynamics is constrained by so-called \emph{$l$-bits}, which have the form of canonical spin-$1/2$ operators and are obtained through unitary transformations on the physical microscopic degrees of freedom \cite{Serbynetal2013, Huseetal2014}. The existence of such $l$-bits has been proven rigorously \cite{Imbrie2016} for a specific class of disordered spin models in 1D, and their importance has been established by several numerical as well as analytical studies \cite{Rosetal2015, RademakerO2016,Youetal2016,InglisP2016,Obrienetal2016,Pekkeretal2017,Goihletal2018,Kulshreshthaetal2018,Mierzejewskietal2018,Pengetal2019}. 

If the suppression of mass/spin transport at long times is a common feature of all localized phases, a fundamental difference emerges at the level of information spreading. Indeed Anderson localization of non-interacting particles implies the absence of any information transfer between local degrees of freedom, associated with the existence of \emph{non-interacting} $l$-bits. On the other hand, MBL exhibits the novel feature of a slow, logarithmic-in-time, growth of entanglement between any subsystem and its complement, which can be explained in terms of \emph{interacting} $l$-bits \cite{Znidaric2008, Bardarson2012a, Nanduri2014, Znidaric2018}. Despite several experimental efforts directed at the characterization of the MBL phase~\cite{M.Schreiber2015, Bordia2015,  Choi2016, Smith2016Many-bodyDisorder, Bordia2017PeriodicallySystem, Luschen2017a, Luschen2017c, Xu2018, Kohlert2019, Rispolietal2019, chiaroetal2020, Guo2020}, a direct evidence of the existence of $l$-bits has so far remained elusive, due to the intrinsic challenges that such a task imposes: namely, high-precision measurements of local observables in different local base. In practical terms, the main evidence of $l$-bit interactions is the long-time dephasing resulting in the slow growth of entanglement, whose observation is a arduous task. 

In a recent study~\cite{Benini2020}, we provided an unambiguous and experimentally accessible signature of $l$-bit interactions for localized systems. We focused on the paradigmatic example of the Heisenberg chain, mappable onto a model of interacting spinless fermions, in a random or quasi-periodic (QP) potential field. There, we explicitly identified the relevant subset of integrals of motion at strong disorders with a collection of oscillating two-levels systems, associated with particles constrained on two physical sites of the lattice. Initializing the dynamics in the $\ket{1010101010\dots}$ Fock state, the existence of such subsystems forming an ensemble of nearly free $l$-bits was found to dominate the short-time dynamics of experimentally relevant quantities; in particular of the logarithmic return probability to the initial state (chosen to be a staggered density arrangement of fermions) — i.e.\ the \emph{Loschmidt echo} (LE) \footnote{Formally, the Loschmidt echo is more generally defined as the scalar product between the evolution of the same state $\ket{\psi_0}$  with two different Hamiltonians, $H_1$ and $H_2$~\cite{Peres1984, Jalabert2001}. The quantity that we denote as LE, also known as \emph{survival probability} or \emph{return probability}, is obtained in the particular case in which one of the two Hamiltonians is set to 0. This slight shift in usage is now customary in the literature of MLB systems.}. The latter quantity displays a striking feature due to the presence of the ensemble of strongly localized $l$-bits, in the form of a periodic sequence of cusp singularities, which are a robust feature in spite of the average over disorder. A simple analytical model, describing the $l$-bits as an ensemble of non-interacting two-level systems (2LS), allows for a quantitative prediction of the appearance of the LE singularities as a result of the Rabi oscillation of the 2LS, as well as of their power-law decay in time \cite{Benini2020}. Periodic cusps of the LE are predicted and observed to be accompanied by periodic dips in the short-time dynamics of the imbalance — namely the population difference between two sub-lattices, which is a very popular marker of localized vs. ergodic dynamics in the system \cite{M.Schreiber2015, Bordia2015}. Furthermore, the $l$-bit interactions, i.e.\ the characteristic signature of MBL dephasing, are shown to bear a striking signature in the dynamics of the LE as well as of the imbalance. The $l$-bits are observed to cross over from an initial regime of decay of their ensemble oscillations ($\propto 1/\sqrt{t}$ with $t$ the time) due to inhomogeneities in the system, to a regime in which the decay is accelerated by the presence of interactions. The crossover to the interaction-induced decay clearly corresponds to the time regime in which the half-system entanglement entropy starts to develop its asymptotic logarithmic growth. In this work, we briefly review the main results of Ref.\ \cite{Benini2020} and show the effective power of the 2LS model to predict the short-time dynamics of relevant observables in the localized regime, such as the spin/particle imbalance. Moreover, we show the results of a different quench protocol, involving the preparation of the system in the ground state of the model at weak and strong disorders, highlighting a deep difference between the two scenarios and exploring the relevance of the initial state preparation.

%%%%%%%%%%%%%%%%%%%%%%%%%%%%%%%%%%%%%%%%%%%%%%%%%%%%%%%%%%%%%%%%%%%%%%%%%%%%%%%%%
\section{Model and protocols}
\label{sec:Model}
%%%%%%%%%%%%%%%%%%%%%%%%%%%%%%%%%%%%%%%%%%%%%%%%%%%%%%%%%%%%%%%%%%%%%%%%%%%%%%%%%
We explore the dynamical properties of a model of fermions with nearest-neighbors interactions in a QP potential~\cite{Iyer2013a, Naldesi2016, Lee2017, Nag2017, Khemani2017, Setiawan2017, zhang2018, Znidaric2018b, Weiner2019, Doggen2019, Shenglong2019}, corresponding to the paradigmatic $S=1/2$ XXZ model \cite{Znidaric2008, Luitz2015a} via Jordan-Wigner mapping \cite{Jordan1928},
\begin{equation}
{\cal H} = \sum_{i=1}^{L-1} \left [ -\frac{J}{2}\left (c_i^\dagger c_{i+1} + {\rm h.c.} \right ) + V n_i n_{i+1} \right ] - \Delta\sum_{i=1}^L h_i n_i,
\label{eq:IAAmodel}
\end{equation}
where $c_i, c_i^\dagger$ and $n_i = c_i^\dagger c_i$ are fermionic operators acting at site $i$, $\Delta$ is the potential amplitude and $h_i=\cos(2\pi \kappa i +\phi)$ with $\kappa$ a suitably irrational number (we choose $\kappa=0.721$ in agreement with Ref.\ \cite{M.Schreiber2015} for comparison) and $\phi$ a \emph{random phase} picked uniformly in the range $[0,2\pi]$. The numerical simulations are carried out via exact diagonalization (ED) techniques \cite{Weinberg2017QuSpin:Chains, Weinberg2018} and the energy scale is set by $J=1$. We restrict to the half-filling sector. For $V=0$, model \eqref{eq:IAAmodel} maps onto the fermionic Aubry-Andr\'e model \cite{Aubry1980}, which is known to show a transition between a delocalized phase and a fully localized one at $\Delta_c = J$, with an energy independent localization length $\xi = 1/\log(\Delta/t)$. In the interacting case, i.e.\ $V = J$, a QP potential of strength $\Delta \gtrsim 4J$  \cite{Naldesi2016} has been found numerically to lead to MBL, with a phase boundary strongly dependent on the energy density, i.e.\ with the emergence of a many-body mobility edge. Modifications of model \eqref{eq:IAAmodel}, with spinful fermions and on-site interactions, have been successfully implemented and its localization properties have been characterized in several experiments involving cold-atoms setups~\cite{M.Schreiber2015, Bordia2015, Bordia2017PeriodicallySystem, Luschen2017c, Rispolietal2019}. 

We explore two different quench protocols: 
(i) We study the unitary evolution under the action of Hamiltonian $\eqref{eq:IAAmodel}$ for different values of the potential strength $\Delta$, starting from an unentangled product state with a charge-density wave profile, i.e.\ $\ket{\psi_0} = \ket{1010101...}$. We monitor the dynamics by computing at each time different local observables such as the logarithmic LE $\lambda(t)$ and the particle imbalance $I(t)$,
\begin{equation}
    \lambda(t) = - \frac{1}{L} \left [ \log |\langle  \psi_0 | e^{-i {\cal H} t} |\psi_0\rangle|^2\right ]_{\rm av}~, \quad 
    I(t) = \frac{1}{L}\sum_i (-1)^i \left( 2\left[\langle n_i \rangle\right]_{\rm av}-1 \right),
\end{equation}
where $[\cdots]_{av}$ indicates a disorder average over multiple QP potential configurations. 

It is worth noting that the logarithmic LE $\lambda(t)$ is currently the subject of considerable interest as the main observable employed in studies of time-dependent dynamical quantum phase transitions~\cite{Heyl2013, Heyl2014, Jurcevic2017, Yang2017, Heyl2018, Guoetal2019, Yin2018, Halimeh2019} (DQPTs). Such dynamical transitions are associated with the appearance of periodic cusps of $\lambda(t)$ at critical times $t^m$, and are believed to represent a new fundamental feature of many-body systems quenched out-of-equilibrium. Nonetheless, caution must be taken in the interpretation of our results in terms of the appearance of DQPTs in localized systems. Indeed, we have shown in \cite{Benini2020} that the LE singularities arising in our results can be fully explained at short times by a model of non-interacting two-level systems, therefore neglecting any relevant many-body effects. 
In protocol (ii), we study a second quench scenario where we prepare the system in the ground-state of \eqref{eq:IAAmodel} at weak ($\Delta_i<1$) or high ($\Delta_i=10$) disorder strengths, and we trigger non trivial time evolution by suddenly changing the value of $\Delta$ to different values $\Delta_f$. Thus, we are able to study the behavior of $\lambda(t)$ starting from two differentinitial states, i.e.\ a Luttinger liquid (LL) state with algebraic correlations for small $\Delta_i$, and a strongly localized state for high $\Delta_i$. We remind ourselves that the ground-state phase diagram of $\eqref{eq:IAAmodel}$ is considerably richer than the fully random potential counterpart, with a transition boundary separating an extended from a localized phase that strongly depends on the interaction $V$ and the disorder strength $\Delta$~\cite{Naldesi2016,Schuster2002}.

%%%%%%%%%%%%%%%%%%%%%%%%%%%%%%%%%%%%%%%%%%%%%%%%%%%%%%%%%%%%%%%%%%%%%%%%%%%%%%%%%
\section{Results}
%%%%%%%%%%%%%%%%%%%%%%%%%%%%%%%%%%%%%%%%%%%%%%%%%%%%%%%%%%%%%%%%%%%%%%%%%%%%%%%%%

%%%%%%%%%%%%%%%%%%%%%%%%%%%%%%%%%%%%%%%%%%%%%%%%%%%%%%%%%%%%%%%%%%%%%%%%%%%%%%%%%
\subsection{Two-site cluster approximation}
\label{sec:2LS}
In Ref.\ \cite{Benini2020}, we have shown that the short-time dynamics of \eqref{eq:IAAmodel} at strong disorder, encoded in observables such as the LE and the imbalance, can be predicted by a model of uncorrelated 2LSs undergoing Rabi oscillations with a frequency $J$. Such effective 2LSs represent an explicit construction framework of the local conserved $l$-bits emerging in the localized regime. This emergent form of integrability represents the most crucial feature of MBL phenomenology, encoded in the general $l$-bit Hamiltonian~\cite{Huseetal2014,Serbyn2013UniversalSystems}
\begin{equation}
    \mathcal{H}_{MBL}=\sum_i K_i \tau_i+\sum_{i>j}K_{ij}\tau_i\tau_j+\sum_{i>j>l}K_{ijl}\tau_i\tau_j\tau_l + \cdots,
    \label{eq:LIOMS}
\end{equation}
where the operators $\tau_i$ are obtained through a series of quasi-local unitary transformations over the microscopic degrees of freedom of the system, e.g.\ $\tau^{\alpha}_i=\hat{U}\sigma^{\alpha}_i\hat{U}^\dagger$ in the case of spins -- with $\alpha=\{x,y,z\}$. The $l$-bits $\tau^{\alpha}_i$ form a complete set of quasi-local integrals of motion, and can be considered a ``dressed" version of the original microscopic operators, with whom they possess a finite spatial overlap~\cite{Bera2015a}. The picture provided by Eq.\ \eqref{eq:LIOMS} proved to be really successful in explaining many features of the MBL phenomenology, most notably the logarithmic growth of entanglement entropies in the strong disorder regime. The slow entanglement generation occurs as each $\tau^{\alpha}_i$-spin precesses around the microscopic magnetic field  generated by all the other $\tau^{\alpha}_i$'s, acquiring a phase that depends on the state of the other $l$-bits. Thus, the dynamics in MBL systems can be summarized in two main stages: (i) in the \emph{short}-time regime the behavior is dominated by the single-spin terms $\sum_i K_i\tau_i$. These do not produce entanglement across the system; (ii) on the other hand, the \emph{long}-time regime is characterized by a slow dephasing due to many-body interactions between distant $l$-bits. This leads to entanglement generation and to the relaxation of physical observables to long-time steady values (although non-thermal ones). The 2LS framework developed in \cite{Benini2020} and reviewed here deals with both dynamical regimes, providing an explicit construction of the $l$-bits based on the microscopic parameters of the Hamiltonian \eqref{eq:IAAmodel} and the choice of the deterministic disorder represented by the QP potential.

In Fig.\ \ref{fig:Fig1} we illustrate a segment of the 1D chain, highlighting the scheme of construction of the 2LSs induced by the spatial profile of the QP potential: the fastest processes in the system dynamics come from those particles that are located on a site $i$ which is nearly resonant with an unoccupied neighbor, where the resonance condition is fulfilled when the hopping is larger than the screened offset in energy between the sites $i, i+1$, i.e.\ $J/2\gtrapprox |\delta_i+V|$, where $\delta_i=h_{i+1}-h_i$. Let focus on a single two-site cluster $(i, i+1)$ where one fermion sits on site $i$ and we assume that site $i+2$ is occupied by another particle while $i-1$ is unoccupied.
{We note that this assumption of having the 2LS as being spatially isolated along the chain is perfectly realistic, given the particular choice of our initial state and the strong anti-correlation of neighboring energy offsets in a QP potential \cite{Benini2020}.} The two-site Hamiltonian can then be written as
\begin{equation}
{\cal H}_{\rm 2S} = -\frac{J}{2} \left ( c^\dagger_i c_{i+1} + c^\dagger_{i+1} c_i \right )  + h_i n_i + (h_{i+1}+V) n_{i+1},    
\end{equation}
and can be translated to the spin language by considering the operator mapping
\begin{equation}
    n_i-n_{i+1}\rightarrow \sigma_i^z, \quad c^\dagger_i c_{i+1} + c^\dagger_{i+1} c_i\rightarrow \sigma_i^x,
\end{equation}
which leads to a simple model of a 2LS with detuning $\delta$ and Rabi frequency $J/2$, i.e.\ 
\begin{equation}
{\cal H}_{\rm 2S} = -\frac{J}{2} \sigma^x + \frac{\delta}{2} \sigma^z + {\rm const.}
\end{equation}
At this point, it is a well-known problem in quantum mechanics to obtain the probability of persistence of the initial configuration in time, starting from the initial state $\ket{10}$ \cite{ScullyBook1997}, namely, one finds
\begin{equation}
 p(\delta;J,t) = \frac{1}{1+(\delta/J)^2} ~\sin^2 \left ( \frac{\sqrt{1+(\delta/J)^2}}{2} tJ \right ),
 \label{eq:pdelta}
\end{equation}
with $\delta=h_{i+1}-h_i -V$.
%%%%%%%%%%%%%%%%%%%%%%%%%%%%%%%%%%%%%%%%%%%%%%%%%%%%%%%%%%%%%%%%%%%%%%%%%%%%%%%%

%%%%%%%%%%%%%%%%%%%%%%%%%%%%%%%%%%%%%%%%%%%%%%%%%%%%%%%%%%%%%%%%%%%%%%%%%%%%%%%%
\subsection{Loschmidt echo and imbalance predictions}
\label{sec:LEIMB}
Starting from the assumption of having a finite number of quasi-resonant 2LSs scattered across the system and surrounded by frozen degrees of freedom,  we can write a tensor product ansatz for the evolved state at time $t$ as
\begin{equation}
    |\psi(t) \rangle \approx \left( \otimes_{n} |\psi^{(n)}_{2LS}(t)\rangle\right) \otimes \left( \otimes'_i |\psi_{0,i}\rangle \right),
\end{equation}
where the first term runs over all the quasi-resonant 2LSs described in \ref{sec:2LS}, labeled by $n$, and the second product includes all the other regions of the chain; see Refs.\ \cite{Sierant2017a,Sierant2017b,Janarek2018} for a similar ansatz to study long-time dynamical properties of MBL systems. Note that in this picture the dynamics comes only from the evolved state of the 2LS regions, while the initial configuration persists in the frozen regions. 
\begin{figure}[tb]
    \centering
    \includegraphics[width=0.8\textwidth]{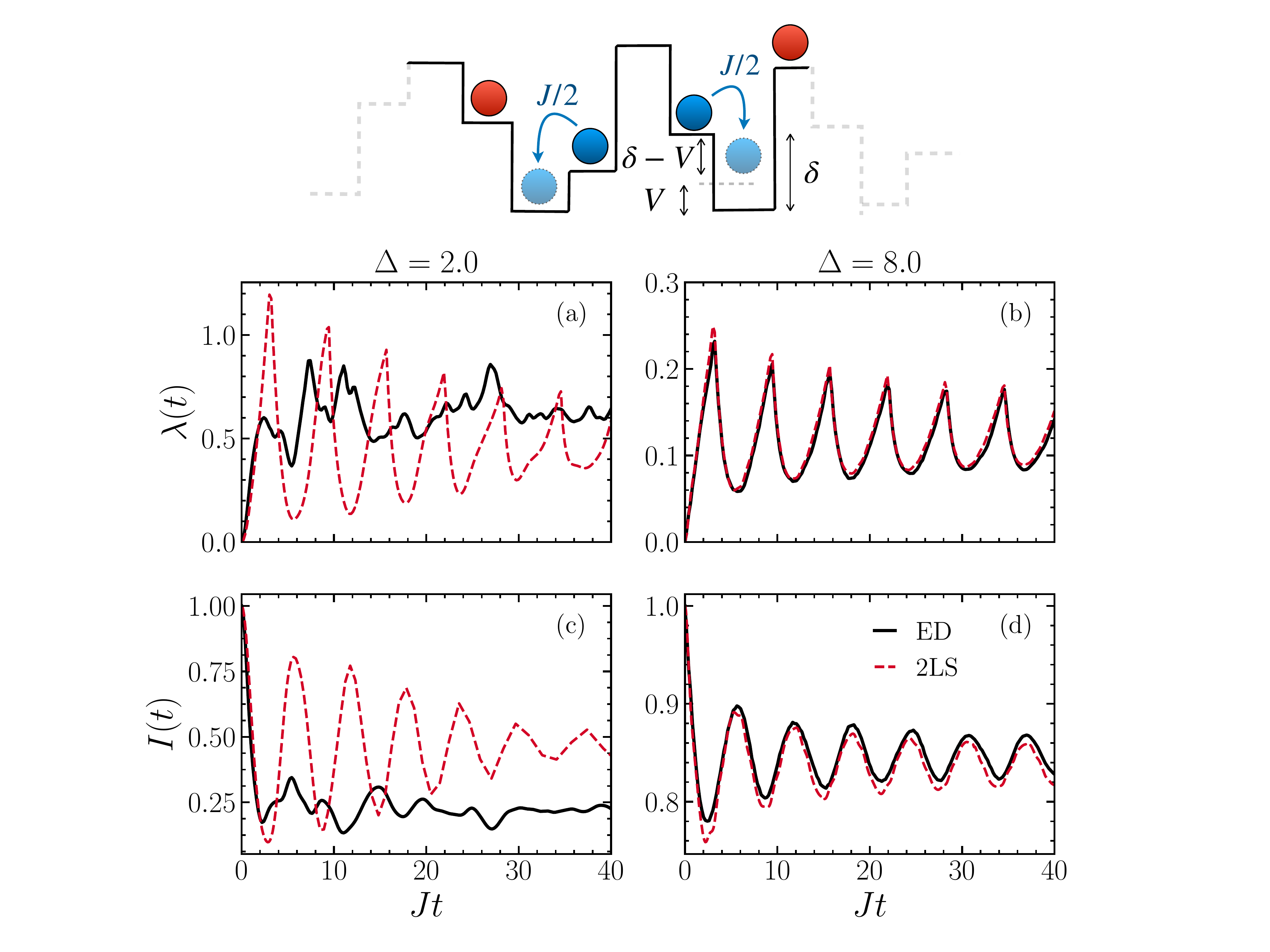} 
    \caption{Top: Sketch of a section of a 1D chain of particles in a in-homogeneous potential. 
    Bottom: Comparison between the numerically obtained $\lambda(t)$ (a-b) and $I(t)$ (c-d) and the corresponding predictions of the 2LS model. The results are shown for two different values of the potential amplitude $\Delta=/{2,8/}$and for a chain of size $L=22$. }
    \label{fig:Fig1}
\end{figure}
From \eqref{eq:pdelta}, it is straightforward to obtain a closed form for the logarithmic LE as
\begin{equation}
    \lambda(t)=-\frac{1}{L}\sum_n \log \left [ 1- p(\delta(n,J,t) \right],
    \label{eq:Rabi1am}
\end{equation}
which upon averaging over disorder reduces to the integral expression
\begin{equation}
    \lambda(t) = -\int P(\delta) \log \left [ 1- p(\delta,J,t) \right]. 
    \label{eq:intlam}
\end{equation}
Here, $P(\delta)$ is the probability distribution of having a detuning between two adjacent sites equal to $\delta$. It depends on the specific form of the external potential and it can be found analytically (as it is the case in our scenario) or sampled numerically; in the case of a QP potential, it is known that \cite{Guarrera2007} 
\begin{equation}
    P(x) =  \frac{1-(x/\tilde{\Delta})^2}{\sqrt{\pi\tilde{\Delta}}}, \quad \text{with } \quad \tilde{\Delta} = \Delta \sin(\pi \kappa).
\end{equation}
Similarly, it is possible to obtain a very simple analytical expression also for the particle imbalance, which is directly related to the LE. Indeed, the imbalance is just given as the probability in \eqref{eq:pdelta}, i.e.\ the persistence of the initial cluster configuration. Thus, the integral form for the imbalance reads
\begin{equation}
    I(t)=\int P(\delta) p(\delta,J,t).
    \label{eq:intimb}
\end{equation}
%%%%%%%%%%%%%%%%%%%%%%%%%%%%%%%%%%%%%%%%%%%%%%%%%%%%%%%%%%%%%%%%%%%%%%%%%%%%%%%%%
\begin{figure}[tb]
    \centering
    \includegraphics[width=0.8\columnwidth]{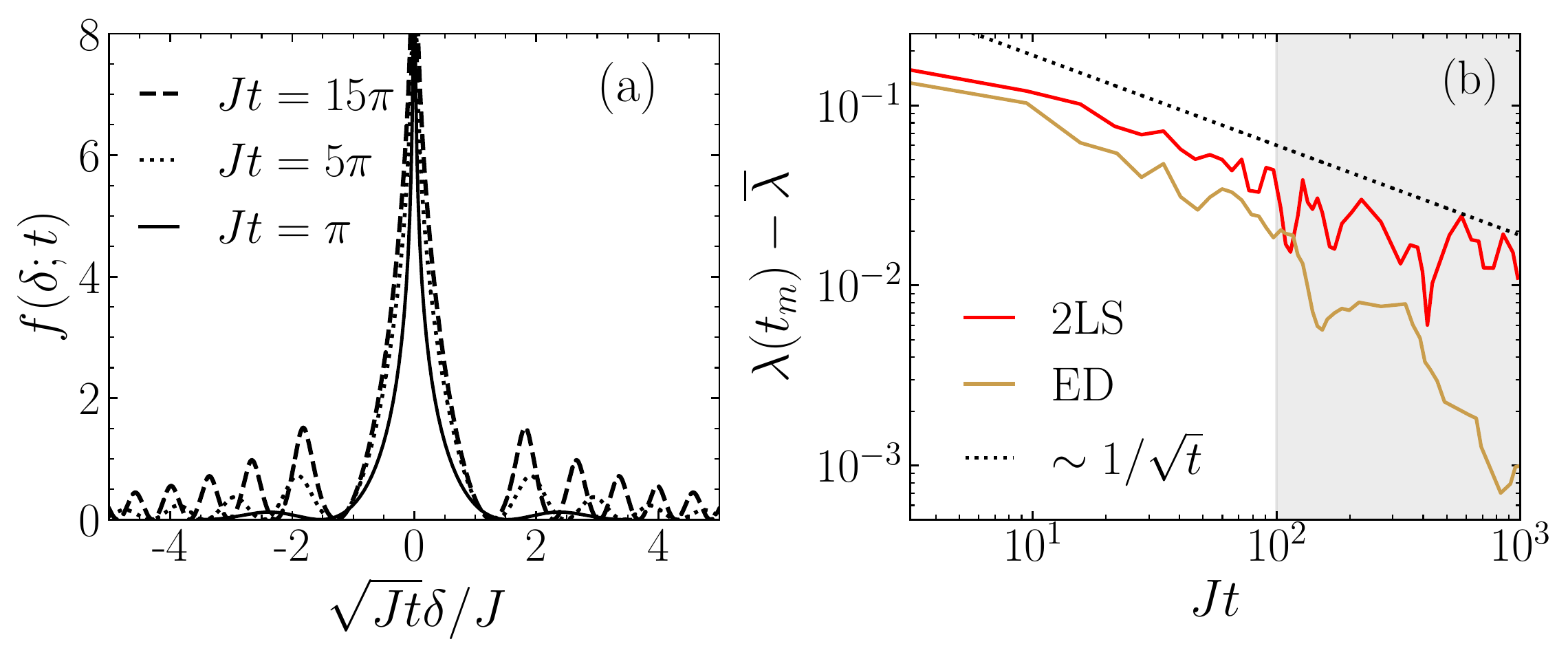}
    \caption{(a) Plot of $f(\delta;t)$ as a function of $\delta$ for different values of the critical times $t_0= \pi/J$, $t_2= 5 \pi/J$ and $t_7=15 \pi/J$. The width of the central peaks becomes independent when plotted against $\sqrt{t}\delta$. (b) Decay of the height of $\lambda(t)$ peaks at long times. The dotted line is a guide to the eye for $1/\sqrt{t}$. Data adapted from \cite{Benini2020}.}
    \label{fig:Fig2}
\end{figure}
%%%%%%%%%%%%%%%%%%%%%%%%%%%%%%%%%%%%%%%%%%%%%%%%%%%%%%%%%%%%%%%%%%%%%%%%%%%%%%%%%

We tested the predictive power of the 2LS model by comparing the numerical solutions of integrals \eqref{eq:intlam} and \eqref{eq:intimb} with ED simulations performed on a chain of $L=22$ sites described by Hamiltonian \eqref{eq:IAAmodel}. The results, illustrated in Fig.\ \ref{fig:Fig1}~($a-d$), show the effective ability of the 2LS model to capture the behavior of both $\lambda(t)$ and $I(t)$ in the strong disorder regime. For $\Delta=8J$, i.e.\ in the MBL phase, the logarithm of the LE displays a periodic sequence of cusps at times $t_m = (2m+1) \pi/J$ $(m=0, 1, 2, ...)$. The emergence of true non-analyticities in the evolution of $\lambda(t)$ is surprising \emph{per se}, given that the singularities survive the disorder-averaging procedure ($\mathcal{O}\sim10^3$ realizations) and are thus robust features of the short-time dynamics of MBL systems. The comparison of the numerical results with the prediction of the 2LS model in Eq.\ \eqref{eq:intlam} shows a convincing agreement, without any need of fine tuning. The peaks in $\lambda(t)$ appear at moments in time when the system is in a configuration maximally different from the initial state, thus signalled by local minima of the imbalance. $I(t)$ starts at its maximal value $I(0)=1$ and decays rapidly to a small value at weak disorder ($\Delta=2J$), while it remains around a finite value deep in the MBL phase at $\Delta=8J$, denoting a persistence of the initial ordering for longer times. Such a behavior is exactly predicted by Eq.\ \eqref{eq:intimb}, which also captures the oscillatory behavior that emerges in the short-time regime. 

Besides providing a simple quantitative model for the prediction of dynamical features of MBL dynamics at strong disorder, an in-depth analysis of the integrand function in \eqref{eq:intlam}, i.e.\ $f(\delta;t)=\log[1-p(\delta,J,t)]$, leads to the origin of the singularities displayed by the LE. In fact, for $\delta$ around zero, i.e.\ close to the effective condition of full resonance in the 2LS description, $f(\delta;t)$ develops a singularity with a width decreasing as $1/\sqrt{t_m}$, as shown in Fig.\ \ref{fig:Fig2}(a). The cusps obtained in the ED simulations (Fig.\ \ref{fig:Fig1}(b-d)) directly come from such a divergent singularity: after odd multiples of half a Rabi oscillation, the system is in a state nearly orthogonal to the initial one. The full divergence becomes a cusp singularity as the set of the fully resonant 2LSs is of zero measure, and the main contribution comes from \emph{quasi}-resonant 2LSs, i.e.\ for $\delta\approx0$. The set of all \emph{quasi}-resonant 2LSs represents an ensemble of approximate \emph{l}-bits that can be described by the first term of \eqref{eq:LIOMS}, i.e.\ the single-bit term $\mathcal{H}=\sum_pK_p\tau_p$, where $p$ labels the pair of sites $(i,i+1)$. The advantage of the 2LS framework is that one can obtain an exact expression for the $l$-bit operators as 
\begin{equation}
    \tau_p=\frac{\delta_p}{K_p}\sigma_p^z-\frac{J}{K_p}\sigma_p^x,
\end{equation}
The $l$-bit coefficients $K_p$ are the 2LS splittings, i.e.\ $K_p=\sqrt{\delta_p^2+J^2}$.

As seen in Fig.\ \ref{fig:Fig1}(b-d) the LE cusp heights decay in time, consistently with the damping of the imbalance oscillations. The height of the cusps is directly determined by the integral of the $f(\delta;t)$ function around $\delta=0$: since the width of the central peak shrinks as $\sqrt{t_m}$, it is safe to assume that the cusp height $\lambda(t_m)-\bar{\lambda}$, as predicted from the 2LS, should also decay as $1/\sqrt{t_m}$. Indeed, this is the case for the 2LS results, as shown in Fig.\ \ref{fig:Fig2}(b). The limits of the 2LS model as a model of \emph{free} independent $l$-bits appear clearly in the comparison with ED data at longer times: the numerical results after $t\gtrsim 100J$ display a strong deviation from the 2LS power-law decay, decreasing in a faster manner. This time-scale sets the onset of the second dynamical regime of MBL systems. The effects of interactions come into play, leading to a faster decay that embodies the dephasing contribution coming from many-body terms of the $l$-bit Hamiltonian \eqref{eq:LIOMS}. In this regime, entanglement production is also expected to include the contribution from $l$-bit interactions, with a logarithmic growth of the entanglement entropy \cite{Bardarson2012a}. The breakdown of the 2LS predictions, marked by a faster decay of the LE cusp heights, allows one to identify a characteristic time $t^*$ which is consistent with the onset of the logarithmic growth of the entanglement entropy \cite{Benini2020}. Furthermore, the same deviation between the 2LS predictions and the ED results can be observed in the damping of the imbalance oscillations, when monitoring the evolution of the minima depths displayed in Fig.\ \ref{fig:Fig1}(d)~\cite{Benini2020}.

%%%%%%%%%%%%%%%%%%%%%%%%%%%%%%%%%%%%%%%%%%%%%%%%%%%%%%%%%%%%%%%%%%%%%%%%%%%%%%%
\subsection{Energy-selective quench}
We present here some results obtained with an alternative quench protocol, briefly described in Sec.\ \ref{sec:Model}. The aim is to provide an energy-resolved method to study dynamical properties of MBL systems through the observation of the LE, via an energy-selective quench \cite{Naldesi2016}. We prepare the system in the ground state of \eqref{eq:IAAmodel} for a given initial potential amplitude $\Delta_i$, and we trigger the quench by suddenly switching the QP potential to a fixed final value $\Delta_f$. The sudden quench injects in the system a controlled amount of energy
\begin{equation}
    \Delta E= \braket{\psi_i|\mathcal{H}(\Delta_f)|\psi_i}-\braket{\psi_f|\mathcal{H}(\Delta_f)|\psi_f},
\end{equation}
where $\ket{\psi_{i(f)}}$ is the ground state of $\mathcal{H}(\Delta_{i(f)})$. Thus, the eigenspectrum of the final Hamiltonian $\mathcal{H}(\Delta_f)$ can be explored by fixing the final disorder strength and continuously varying $\Delta_i$. A wider quench in the disorder amplitude leads to eigenstates of the final Hamiltonian with parametrically higher energy density $\epsilon=\Delta E/W$, where $W$ is the spectral width of $\mathcal{H}(\Delta_f)$. 

At the ground-state level, model \eqref{eq:IAAmodel} is characterized by a quantum critical point separating a gapless Luttinger liquid phase and a Bose-glass phase, the latter displaying exponentially decaying correlations~\cite{Naldesi2016}. The critical disorder at which the system undergoes a ground state localization transition in our case is at $\Delta_c\simeq 2.5J$. With respect to the non-interacting Aubry-André case ($V=0$), the critical point at $\Delta_c=J$ is shifted towards higher values of the disorder due to the screening effect of repulsive interactions for $V=J$. Above the ground state, the full eigenspectrum exhibits a many-body mobility edge (MBME), between an ergodic phase and an MBL phase. For disorder strengths in the interval $1.5J \lesssim \Delta \lesssim 4J$, the system is characterized by the coexistence of extended and localized states depending on the energy density.
%%%%%%%%%%%%%%%%%%%%%%%%%%%%%%%%%%%%%%%%%%%%%%%%%%%%%%%%%%%%%%%%%%%%%%%%%%%%%%%%
\begin{figure}[tb]
    \centering
    \includegraphics[width=0.6\textwidth]{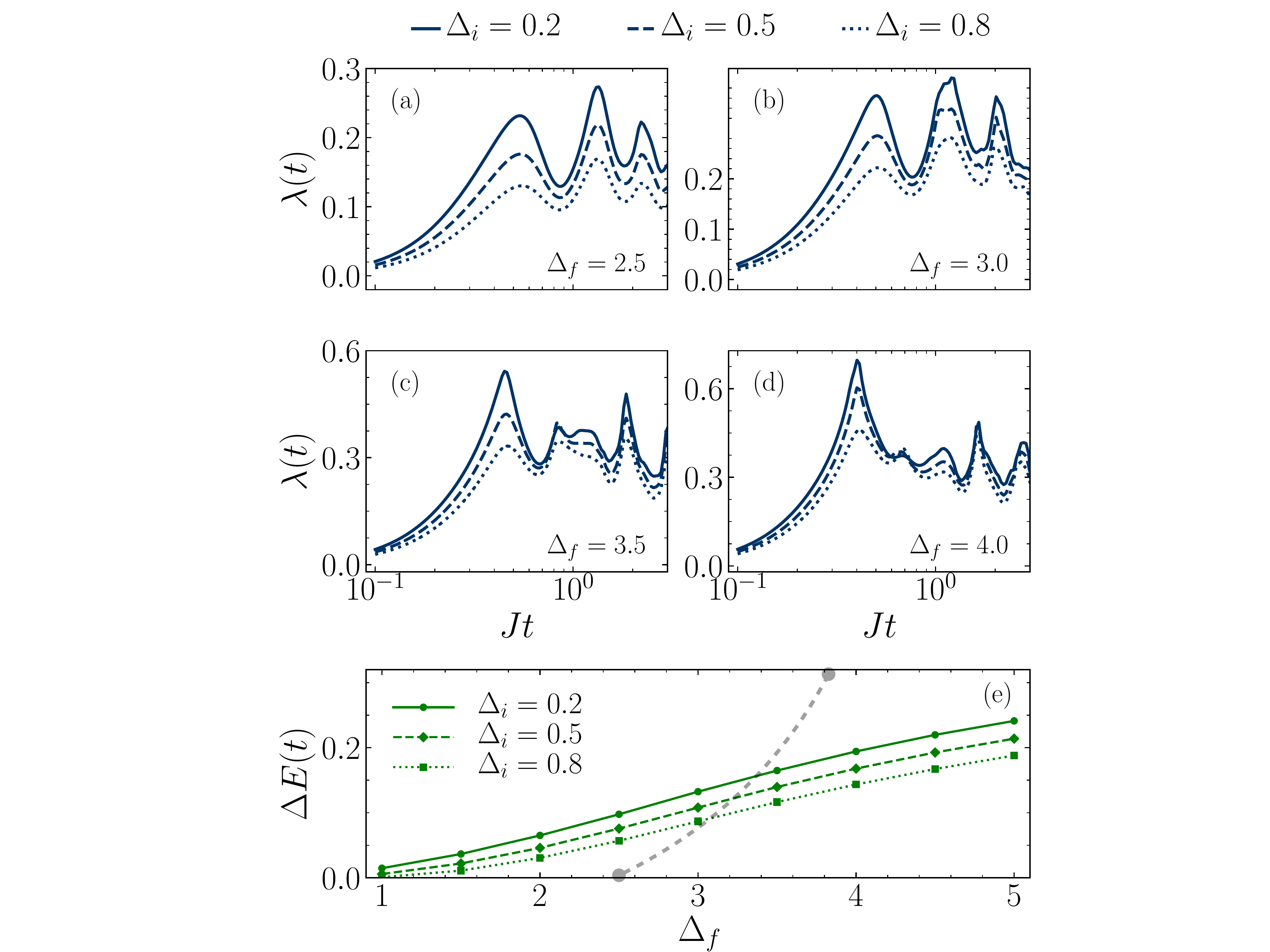} 
    \caption{(a-d) Loschmidt return rate dynamics for the ground-state of Hamiltonian \eqref{eq:IAAmodel} for different initial QP potential strengths $\Delta_i= 0.2, 0.5, 0.8 $. The quench is realized at fixed final disorder values $\Delta_f= 2.5, 3.0, 3.5, 4.0$, in proximity of the critical region separating extended from MBL phases. (e) Energy injected through the quench for different initial $\Delta_i$ as a function of $\Delta_f$. The grey dashed line is a guide to the eye depicting the MBME at the relevant energy densities, from data adapted from \cite{Naldesi2016}.}
    \label{fig:Fig4}
\end{figure}
% %%%%%%%%%%%%%%%%%%%%%%%%%%%%%%%%%%%%%%%%%%%%%%%%%%%%%%%%%%%%%%%%%%%%%%%%%%%%%%%%%

We study the behavior of the Loschmidt return rate $\lambda(t)$ following the quench-spectroscopy scheme described above. When preparing the system in the ground-state of $ \mathcal{H}(\Delta_i)$ and quenching to final disorder strengths $\Delta_f>\Delta_i$, we monitor the out-of-equilibrium dynamics through the computation of $\lambda(t)$ and $\Delta E$. The scope of the investigation is two-fold: does the LE dynamics provide any robust signature of the MBL transition? And is that signature able to detect the MBME? Figs.\ \ref{fig:Fig4}(a-d) illustrate the ED results of the quench obtained for three different values of the initial disorder $\Delta_i= 0.2, 0.5, 0.8$ at four fixed values of the quenching disorder $\Delta_f= 2.5, 3.0 ,3.5, 4.0$, in a chain of $L=20$ sites. The signature revealed by the LE is again the emergence of a sharp singularity in the short-time dynamics as we cross the critical disorder $\Delta_c$ and we enter into the MBL regime. The critical region identified by the onset of the first non-analyticities is $\Delta_f \in [3.5, 4.0]$ which is consistent with the values predicted by the MBME at the energy density reached through our quenches, i.e.\ $\epsilon \sim 0.2$ \cite{Naldesi2016}. 
As can be seen in Fig.\ \ref{fig:Fig4}(e), the energy injected in the system at different $\Delta_f$ increases with the quench width separating $\Delta_i$ and $\Delta_f$. At this stage, the qualitative analysis of the Loschmidt singularities does not provide a confirmation of the existence of an MBME. The shrinking of the peaks of $\lambda(t)$ leading to true singularities seems to be largely independent of the choice of $\Delta_i$, and thus of the energy density injected through the quench. The non-analytic behavior emerges more evidently for \emph{large} quenches, i.e.\ for increasing $\Delta_f-\Delta_i$. Intuitively, this is consistent with the interpretation of $\lambda(t)$ singularities as minima of the survival probability: as the distance $\Delta_f - \Delta_i$ increases, the evolution driven by $\mathcal{H}_f$ brings the evolved state $\ket{\psi(t)}$ to be farther from the ground-state of $\mathcal{H}_i$ more rapidly. On the other hand, the energy range explored through this series of quenches is too restricted for a detailed study of the shape of the MBME.

\begin{figure}[tb]
    \centering
    \includegraphics[width=0.42\textwidth]{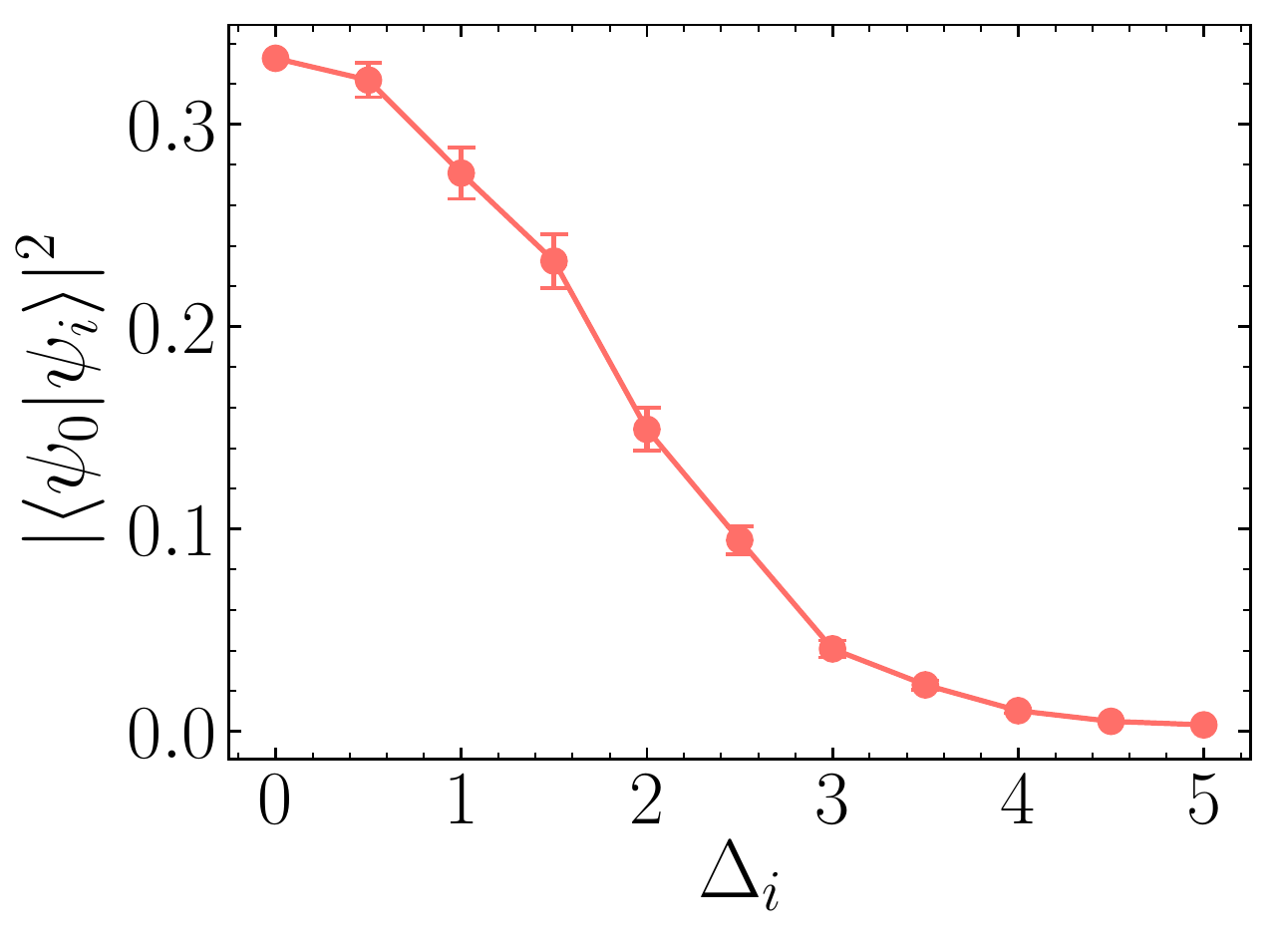} 
    \caption{Quantum overlap between the CDW state $\ket{\psi_0} = \ket{1010101...}$ and the ground-state $\ket{\psi_i}$ of \eqref{eq:IAAmodel} as a function of the initial potential strength $\Delta_i$ for $L=16$. }
    \label{fig:Fig5}
\end{figure}

The 2LS framework described in Sec.\ref{sec:2LS} offers a qualitative explanation for the emergence of singularities in the dynamics of the LE, in spite of the fact that the initial state for the present quenches is at first sight very different from that considered in previous sections. We recall that an essential ingredient underlying the construction of the 2LS model is the single particle occupation of each well of the QP potential (whose quasi-period is between three and four sites). This is clearly satisfied by the choice of a CDW-like factorized state, which is precisely the case studied in Sec.~\ref{sec:LEIMB}. As $\Delta_i\rightarrow 0$, the ground state of model \eqref{eq:IAAmodel} develops very staggered density-density correlations decaying algebraically with the distance, with the main contributions coming equally from the basis states $\ket{\psi_0}=\ket{101010\dots}$ and $\overline{\ket{\psi_0}}=\ket{010101\dots}$. Indeed, as can be seen in Fig.\ \ref{fig:Fig5}, the overlap between the ground-state $\ket{\psi_i}$ and the CDW state $\ket{\psi_0} = \ket{1010101...}$ is significant for very low $\Delta_i$, being maximal in the extreme case $\Delta_i=0$. Moreover, the overlap with $\overline{\ket{\psi_0}}$ provides the same quantitative contribution (not shown). Hence, the similarity of the initial ground state for small $\Delta_i$ with $\ket{\psi_0}$ and $\overline{\ket{\psi_0}}$ explains the appearance of singularities in the LE dynamics, albeit less pronounced than in the pure CDW case. 

On the other hand, $|\braket{\psi_0|\psi_i}|^2$ has a fast decay as we increase $\Delta_i$, becoming negligible for $\Delta_i\gtrsim 3$. An intuitive expectation is that evolving an initial ground state already localized by a strong QP potential, i.e.\ for high $\Delta_i$, would not lead to singular behavior. This is indeed confirmed by the results presented in Fig.\ref{fig:Fig5}, in which we show a comparison between different choices of the initial ground-state. With a quench to QP potential strengths $\Delta_f \in [1.0, 8.0]$, one observes the emergence of singularities for $\Delta_i=0.8$ as we cross the ETH-MBL transition point, as discussed above. Moreover, the peaks become disorder-independent when the time axis is rescaled as $1/t_c$, where $t_c$ is defined as the first relative maximum of $\lambda(t)$. Conversely, the LE evolution of the ground state deep in the localized phase with $\Delta_i=10$ displays a smooth behavior for all values of $\Delta_f$. 

%%%%%%%%%%%%%%%%%%%%%%%%%%%%%%%%%%%%%%%%%%%%%%%%%%%%%%%%%%%%%%%%%%%%%%%%%%%%%%%%%
\section{Conclusions}
%%%%%%%%%%%%%%%%%%%%%%%%%%%%%%%%%%%%%%%%%%%%%%%%%%%%%%%%%%%%%%%%%%%%%%%%%%%%%%%%%
In this paper, we described the LE dynamics in many-body localized systems for two quench scenarios.
In the first scenario, we studied the periodic structure of cusp-like singularities emerging in the LE during the unitary evolution governed by the Hamiltonian of interacting spinless fermions in a strong QP potential, and initialized in an unentangled product state with CDW order. We discussed a simple model of approximate $l$-bits, explicitly identified in the strong MBL regime as a collection of independently oscillating 2LSs. The singularities in the LE emerge naturally from the study of the resonance condition between the $l$-bits, leading to very accurate predictions, in the short-time limit, of the behavior of the LE, as well as of other observables such as the particle imbalance. The deviation from the 2LS model predictions offers as well a way to identify the dynamical regime where the dephasing induced by the many-body interactions comes into play, thus marking the onset of true MBL dynamics. 

In a second quench scenario, we moved closer to the delicate MBL-transition regime, exploring the out-of-equilibrium dynamics with a quench spectroscopy protocol. Initializing the system in an extended initial state, i.e.\ the ground-state of Hamiltonian \eqref{eq:IAAmodel} at low disorder, and triggering time evolution by a sudden switch of the disorder strength to $\Delta_f>\Delta_i$, we observed the shrinking of LE peaks into true singularities at post-quench disorder values compatible with previous estimates of the critical disorder ($\Delta > \Delta_c\sim3.5-4.0$). The explanation for the appearance of such singular behavior comes from the significant overlap of the ground state of the model at weak disorders with the factorized CDW state employed in the first quench scenario. This suggests that the quench dynamics of a ground state in the extended phase can be again understood in the framework of the 2LS model, even at intermediate disorder regimes.

The interplay of many-body interactions and disorder has long been a challenge in condensed matter physics \cite{Anderson-book,Belitz1994}. With the advent of quantum simulators, able to mimic disordered and interacting quantum many-body systems, a new tool has been given to researchers for controlled and quantitative experiments. However, this calls for the development of reliable theoretical concepts to measure and characterize such systems. Our work provides a conceptually simple framework, in the form of the 2LS model, in which one can both reach a fundamental understanding of the non-equilibrium dynamics of MBL systems and obtain reliable quantitative predictions for the behavior of experimentally accessible observables.

\begin{figure}[tb]
    \centering
    \includegraphics[width=0.8\textwidth]{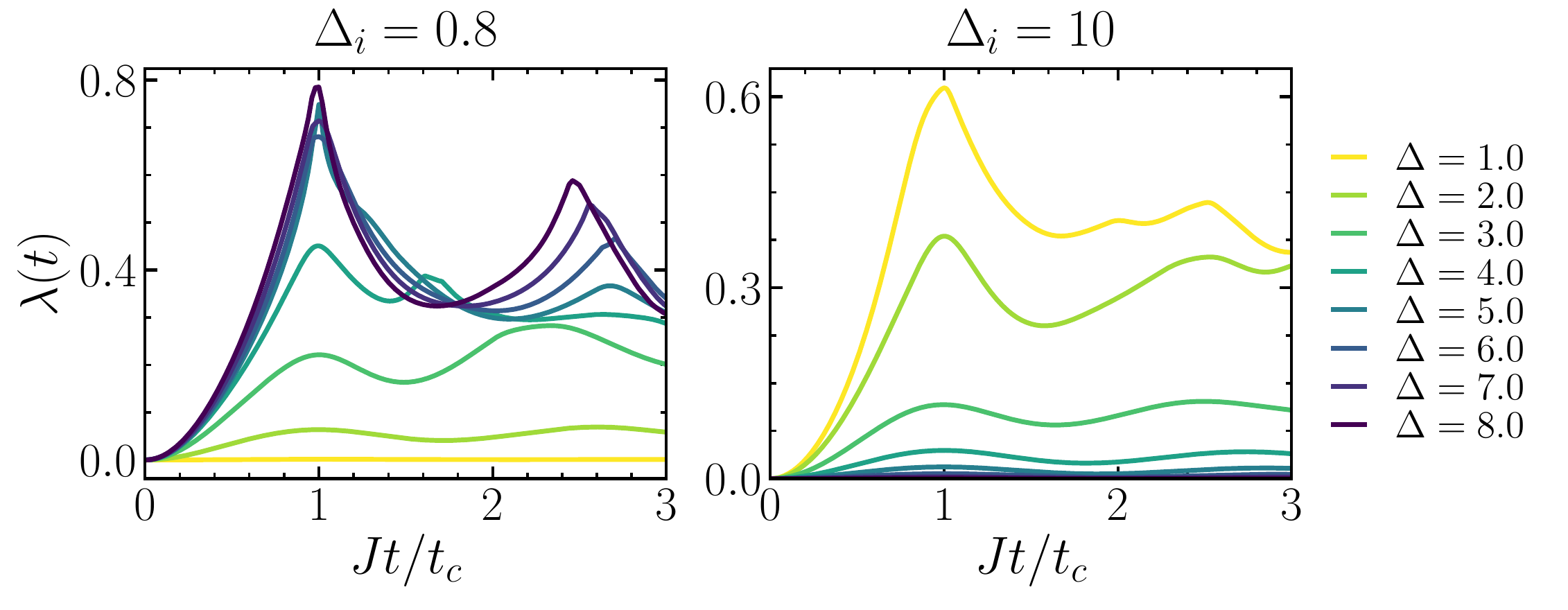} 
    \caption{Quench dynamics of $\lambda(t)$ starting from a low initial disorder $\Delta_i=0.8$ (left) and a high initial disorder $\Delta_i=10$. The time axis has been rescaled as $1/t_c$ to obtain disorder independence.}
    \label{fig:Fig3}
\end{figure}

\section*{Acknowledgements}
L.B.\ gratefully acknowledges hospitality and financial support from the Labora\-toi\-re de Physi\-que of the ENS Lyon. We thank Warwick's Scientific Computing Research Technology Platform and HPC Midlands+ (Athena) for computing time and support (EPSRC on grant EP/P020232/1). Most of the numerical simulations were carried out using routines contained in the \emph{QuSpin}~\cite{Weinberg2017QuSpin:Chains, Weinberg2018} open-source library. 
UK data statement: Data supporting this work is available as detailed in Ref.\ \cite{Benini2020}.

\bibliographystyle{elsarticle-num}
\bibliography{references.bib}

\begin{thebibliography}{10}
\expandafter\ifx\csname url\endcsname\relax
  \def\url#1{\texttt{#1}}\fi
\expandafter\ifx\csname urlprefix\endcsname\relax\def\urlprefix{URL }\fi
\expandafter\ifx\csname href\endcsname\relax
  \def\href#1#2{#2} \def\path#1{#1}\fi

\bibitem{Akkermans-book}
E.~Akkermans, G.~Montambaux, Mesoscopic Physics of Electrons and Photons,
  Cambridge University Press, 2007.

\bibitem{Anderson-book}
T.~Brandes, S.~Ketteman (Eds.), Anderson Localization and its Ramifications,
  Springer, Berlin, 2003.

\bibitem{Anderson1958AbsenceLattices}
P.~W. Anderson, {Absence of diffusion in certain random lattices}, Physical
  Review 109~(5) (1958) 1492--1505.
\newblock \href {https://doi.org/10.1103/PhysRev.109.1492}
  {\path{doi:10.1103/PhysRev.109.1492}}.

\bibitem{NandkishoreH2015}
R.~Nandkishore, D.~A. Huse,
  \href{https://doi.org/10.1146/annurev-conmatphys-031214-014726}{Many-body
  localization and thermalization in quantum statistical mechanics}, Annual
  Review of Condensed Matter Physics 6~(1) (2015) 15--38.
\newblock \href
  {http://arxiv.org/abs/https://doi.org/10.1146/annurev-conmatphys-031214-014726}
  {\path{arXiv:https://doi.org/10.1146/annurev-conmatphys-031214-014726}},
  \href {https://doi.org/10.1146/annurev-conmatphys-031214-014726}
  {\path{doi:10.1146/annurev-conmatphys-031214-014726}}.
\newline\urlprefix\url{https://doi.org/10.1146/annurev-conmatphys-031214-014726}

\bibitem{Alet2018}
F.~Alet, N.~Laflorencie,
  \href{https://www.sciencedirect.com/science/article/pii/S163107051830032X?via%3Dihub}{{Many-body
  localization: An introduction and selected topics}}, Comptes Rendus Physique
  19~(6) (2018) 498--525.
\newblock \href {https://doi.org/10.1016/j.crhy.2018.03.003}
  {\path{doi:10.1016/j.crhy.2018.03.003}}.
\newline\urlprefix\url{https://www.sciencedirect.com/science/article/pii/S163107051830032X?via%3Dihub}

\bibitem{Abaninetal2019}
D.~A. Abanin, E.~Altman, I.~Bloch, M.~Serbyn,
  \href{https://link.aps.org/doi/10.1103/RevModPhys.91.021001}{Colloquium:
  Many-body localization, thermalization, and entanglement}, Rev. Mod. Phys. 91
  (2019) 021001.
\newblock \href {https://doi.org/10.1103/RevModPhys.91.021001}
  {\path{doi:10.1103/RevModPhys.91.021001}}.
\newline\urlprefix\url{https://link.aps.org/doi/10.1103/RevModPhys.91.021001}

\bibitem{GOPALAKRISHNAN2020}
S.~Gopalakrishnan, S.~Parameswaran,
  \href{http://www.sciencedirect.com/science/article/pii/S0370157320301083}{Dynamics
  and transport at the threshold of many-body localization}, Physics Reports
  862 (2020) 1 -- 62, dynamics and transport at the threshold of many-body
  localization.
\newblock \href {https://doi.org/https://doi.org/10.1016/j.physrep.2020.03.003}
  {\path{doi:https://doi.org/10.1016/j.physrep.2020.03.003}}.
\newline\urlprefix\url{http://www.sciencedirect.com/science/article/pii/S0370157320301083}

\bibitem{Serbynetal2013}
M.~Serbyn, Z.~Papi\ifmmode~\acute{c}\else \'{c}\fi{}, D.~A. Abanin,
  \href{https://link.aps.org/doi/10.1103/PhysRevLett.111.127201}{Local
  conservation laws and the structure of the many-body localized states}, Phys.
  Rev. Lett. 111 (2013) 127201.
\newblock \href {https://doi.org/10.1103/PhysRevLett.111.127201}
  {\path{doi:10.1103/PhysRevLett.111.127201}}.
\newline\urlprefix\url{https://link.aps.org/doi/10.1103/PhysRevLett.111.127201}

\bibitem{Huseetal2014}
D.~A. Huse, R.~Nandkishore, V.~Oganesyan,
  \href{https://link.aps.org/doi/10.1103/PhysRevB.90.174202}{Phenomenology of
  fully many-body-localized systems}, Phys. Rev. B 90 (2014) 174202.
\newblock \href {https://doi.org/10.1103/PhysRevB.90.174202}
  {\path{doi:10.1103/PhysRevB.90.174202}}.
\newline\urlprefix\url{https://link.aps.org/doi/10.1103/PhysRevB.90.174202}

\bibitem{Imbrie2016}
J.~Z. Imbrie,
  \href{https://link.aps.org/doi/10.1103/PhysRevLett.117.027201}{Diagonalization
  and many-body localization for a disordered quantum spin chain}, Phys. Rev.
  Lett. 117 (2016) 027201.
\newblock \href {https://doi.org/10.1103/PhysRevLett.117.027201}
  {\path{doi:10.1103/PhysRevLett.117.027201}}.
\newline\urlprefix\url{https://link.aps.org/doi/10.1103/PhysRevLett.117.027201}

\bibitem{Imbrieetal2017}
J.~Z. Imbrie, V.~Ros, A.~Scardicchio, Local integrals of motion in many-body
  localized systems, Annalen der Physik 529~(7) (2017) 1600278.
\newblock \href {https://doi.org/10.1002/andp.201600278}
  {\path{doi:10.1002/andp.201600278}}.

\bibitem{Rosetal2015}
V.~Ros, M.~M\"uller, A.~Scardicchio,
  \href{http://www.sciencedirect.com/science/article/pii/S0550321314003836}{Integrals
  of motion in the many-body localized phase}, Nuclear Physics B 891 (2015) 420
  -- 465.
\newblock \href
  {https://doi.org/https://doi.org/10.1016/j.nuclphysb.2014.12.014}
  {\path{doi:https://doi.org/10.1016/j.nuclphysb.2014.12.014}}.
\newline\urlprefix\url{http://www.sciencedirect.com/science/article/pii/S0550321314003836}

\bibitem{RademakerO2016}
L.~Rademaker, M.~Ortu\~no,
  \href{https://link.aps.org/doi/10.1103/PhysRevLett.116.010404}{Explicit local
  integrals of motion for the many-body localized state}, Phys. Rev. Lett. 116
  (2016) 010404.
\newblock \href {https://doi.org/10.1103/PhysRevLett.116.010404}
  {\path{doi:10.1103/PhysRevLett.116.010404}}.
\newline\urlprefix\url{https://link.aps.org/doi/10.1103/PhysRevLett.116.010404}

\bibitem{Youetal2016}
Y.-Z. You, X.-L. Qi, C.~Xu,
  \href{https://link.aps.org/doi/10.1103/PhysRevB.93.104205}{Entanglement
  holographic mapping of many-body localized system by spectrum bifurcation
  renormalization group}, Phys. Rev. B 93 (2016) 104205.
\newblock \href {https://doi.org/10.1103/PhysRevB.93.104205}
  {\path{doi:10.1103/PhysRevB.93.104205}}.
\newline\urlprefix\url{https://link.aps.org/doi/10.1103/PhysRevB.93.104205}

\bibitem{InglisP2016}
S.~Inglis, L.~Pollet,
  \href{https://link.aps.org/doi/10.1103/PhysRevLett.117.120402}{Accessing
  many-body localized states through the generalized gibbs ensemble}, Phys.
  Rev. Lett. 117 (2016) 120402.
\newblock \href {https://doi.org/10.1103/PhysRevLett.117.120402}
  {\path{doi:10.1103/PhysRevLett.117.120402}}.
\newline\urlprefix\url{https://link.aps.org/doi/10.1103/PhysRevLett.117.120402}

\bibitem{Obrienetal2016}
T.~E. O'Brien, D.~A. Abanin, G.~Vidal, Z.~Papi\ifmmode~\acute{c}\else
  \'{c}\fi{},
  \href{https://link.aps.org/doi/10.1103/PhysRevB.94.144208}{Explicit
  construction of local conserved operators in disordered many-body systems},
  Phys. Rev. B 94 (2016) 144208.
\newblock \href {https://doi.org/10.1103/PhysRevB.94.144208}
  {\path{doi:10.1103/PhysRevB.94.144208}}.
\newline\urlprefix\url{https://link.aps.org/doi/10.1103/PhysRevB.94.144208}

\bibitem{Pekkeretal2017}
D.~Pekker, B.~K. Clark, V.~Oganesyan, G.~Refael,
  \href{https://link.aps.org/doi/10.1103/PhysRevLett.119.075701}{Fixed points
  of wegner-wilson flows and many-body localization}, Phys. Rev. Lett. 119
  (2017) 075701.
\newblock \href {https://doi.org/10.1103/PhysRevLett.119.075701}
  {\path{doi:10.1103/PhysRevLett.119.075701}}.
\newline\urlprefix\url{https://link.aps.org/doi/10.1103/PhysRevLett.119.075701}

\bibitem{Goihletal2018}
M.~Goihl, M.~Gluza, C.~Krumnow, J.~Eisert,
  \href{https://link.aps.org/doi/10.1103/PhysRevB.97.134202}{Construction of
  exact constants of motion and effective models for many-body localized
  systems}, Phys. Rev. B 97 (2018) 134202.
\newblock \href {https://doi.org/10.1103/PhysRevB.97.134202}
  {\path{doi:10.1103/PhysRevB.97.134202}}.
\newline\urlprefix\url{https://link.aps.org/doi/10.1103/PhysRevB.97.134202}

\bibitem{Kulshreshthaetal2018}
A.~K. Kulshreshtha, A.~Pal, T.~B. Wahl, S.~H. Simon,
  \href{https://link.aps.org/doi/10.1103/PhysRevB.98.184201}{Behavior of l-bits
  near the many-body localization transition}, Phys. Rev. B 98 (2018) 184201.
\newblock \href {https://doi.org/10.1103/PhysRevB.98.184201}
  {\path{doi:10.1103/PhysRevB.98.184201}}.
\newline\urlprefix\url{https://link.aps.org/doi/10.1103/PhysRevB.98.184201}

\bibitem{Mierzejewskietal2018}
M.~Mierzejewski, M.~Kozarzewski, P.~Prelov\ifmmode~\check{s}\else \v{s}\fi{}ek,
  \href{https://link.aps.org/doi/10.1103/PhysRevB.97.064204}{Counting local
  integrals of motion in disordered spinless-fermion and hubbard chains}, Phys.
  Rev. B 97 (2018) 064204.
\newblock \href {https://doi.org/10.1103/PhysRevB.97.064204}
  {\path{doi:10.1103/PhysRevB.97.064204}}.
\newline\urlprefix\url{https://link.aps.org/doi/10.1103/PhysRevB.97.064204}

\bibitem{Pengetal2019}
P.~Peng, Z.~Li, H.~Yan, K.~X. Wei, P.~Cappellaro,
  \href{https://link.aps.org/doi/10.1103/PhysRevB.100.214203}{Comparing
  many-body localization lengths via nonperturbative construction of local
  integrals of motion}, Phys. Rev. B 100 (2019) 214203.
\newblock \href {https://doi.org/10.1103/PhysRevB.100.214203}
  {\path{doi:10.1103/PhysRevB.100.214203}}.
\newline\urlprefix\url{https://link.aps.org/doi/10.1103/PhysRevB.100.214203}

\bibitem{Znidaric2008}
M.~{\v{Z}}nidari{\v{c}}, T.~Prosen, P.~Prelov{\v{s}}ek,
  \href{https://0-journals-aps-org.pugwash.lib.warwick.ac.uk/prb/pdf/10.1103/PhysRevB.77.064426}{{Many-body
  localization in the Heisenberg XXZ magnet in a random field}}, Physical
  Review B 77~(6) (2008).
\newblock \href {https://doi.org/10.1103/PhysRevB.77.064426}
  {\path{doi:10.1103/PhysRevB.77.064426}}.
\newline\urlprefix\url{https://0-journals-aps-org.pugwash.lib.warwick.ac.uk/prb/pdf/10.1103/PhysRevB.77.064426}

\bibitem{Bardarson2012a}
J.~H. Bardarson, F.~Pollmann, J.~E. Moore,
  \href{https://0-journals-aps-org.pugwash.lib.warwick.ac.uk/prl/pdf/10.1103/PhysRevLett.109.017202}{{Unbounded
  growth of entanglement in models of many-body localization}}, Physical Review
  Letters 109~(1) (2012).
\newblock \href {https://doi.org/10.1103/PhysRevLett.109.017202}
  {\path{doi:10.1103/PhysRevLett.109.017202}}.
\newline\urlprefix\url{https://0-journals-aps-org.pugwash.lib.warwick.ac.uk/prl/pdf/10.1103/PhysRevLett.109.017202}

\bibitem{Nanduri2014}
A.~Nanduri, H.~Kim, D.~A. Huse,
  \href{https://link.aps.org/doi/10.1103/PhysRevB.90.064201}{Entanglement
  spreading in a many-body localized system}, Phys. Rev. B 90 (2014) 064201.
\newblock \href {https://doi.org/10.1103/PhysRevB.90.064201}
  {\path{doi:10.1103/PhysRevB.90.064201}}.
\newline\urlprefix\url{https://link.aps.org/doi/10.1103/PhysRevB.90.064201}

\bibitem{Znidaric2018}
M.~\ifmmode \check{Z}\else \v{Z}\fi{}nidari\ifmmode~\check{c}\else \v{c}\fi{},
  \href{https://link.aps.org/doi/10.1103/PhysRevB.97.214202}{Entanglement in a
  dephasing model and many-body localization}, Phys. Rev. B 97 (2018) 214202.
\newblock \href {https://doi.org/10.1103/PhysRevB.97.214202}
  {\path{doi:10.1103/PhysRevB.97.214202}}.
\newline\urlprefix\url{https://link.aps.org/doi/10.1103/PhysRevB.97.214202}

\bibitem{M.Schreiber2015}
{M. Schreiber}, {S. S. Hodgman}, {P. Bordia}, {H. P. L{\"{u}}schen}, {M. H.
  Fischer}, {R. Vosk}, E.~Altman, {U. Schneider}, I.~Bloch,
  \href{http://www.sciencemag.org/content/349/6250/842.abstract}{{Observation
  of many-body localization of interacting fermions in a quasirandom optical
  lattice}}, Science 349~(6250) (2015) 842--845.
\newblock \href {https://doi.org/10.1126/science.aaa7432}
  {\path{doi:10.1126/science.aaa7432}}.
\newline\urlprefix\url{http://www.sciencemag.org/content/349/6250/842.abstract}

\bibitem{Bordia2015}
P.~Bordia, H.~P. L{\"{u}}schen, S.~S. Hodgman, M.~Schreiber, I.~Bloch,
  U.~Schneider,
  \href{http://arxiv.org/abs/1509.00478%0Ahttp://dx.doi.org/10.1103/PhysRevLett.116.140401}{{Coupling
  Identical 1D Many-Body Localized Systems}}, Physical Review Letters 116~(14)
  (2016) 140401.
\newblock \href {https://doi.org/10.1103/PhysRevLett.116.140401}
  {\path{doi:10.1103/PhysRevLett.116.140401}}.
\newline\urlprefix\url{http://arxiv.org/abs/1509.00478%0Ahttp://dx.doi.org/10.1103/PhysRevLett.116.140401}

\bibitem{Choi2016}
J.-y. Choi, S.~Hild, J.~Zeiher, P.~Schau{\ss}, A.~Rubio-Abadal, T.~Yefsah,
  V.~Khemani, D.~A. Huse, I.~Bloch, C.~Gross,
  \href{https://science.sciencemag.org/content/352/6293/1547}{Exploring the
  many-body localization transition in two dimensions}, Science 352~(6293)
  (2016) 1547--1552.
\newblock \href
  {http://arxiv.org/abs/https://science.sciencemag.org/content/352/6293/1547.full.pdf}
  {\path{arXiv:https://science.sciencemag.org/content/352/6293/1547.full.pdf}},
  \href {https://doi.org/10.1126/science.aaf8834}
  {\path{doi:10.1126/science.aaf8834}}.
\newline\urlprefix\url{https://science.sciencemag.org/content/352/6293/1547}

\bibitem{Smith2016Many-bodyDisorder}
J.~Smith, A.~Lee, P.~Richerme, B.~Neyenhuis, P.~W. Hess, P.~Hauke, M.~Heyl,
  D.~A. Huse, C.~Monroe, \href{www.nature.com/naturephysics}{{Many-body
  localization in a quantum simulator with programmable random disorder}},
  Nature Physics 12~(10) (2016) 907--911.
\newblock \href {https://doi.org/10.1038/nphys3783}
  {\path{doi:10.1038/nphys3783}}.
\newline\urlprefix\url{www.nature.com/naturephysics}

\bibitem{Bordia2017PeriodicallySystem}
P.~Bordia, H.~L{\"{u}}schen, U.~Schneider, M.~Knap, I.~Bloch,
  \href{www.nature.com/naturephysics}{{Periodically driving a many-body
  localized quantum system}}, Nature Physics 13~(5) (2017) 460--464.
\newblock \href {https://doi.org/10.1038/nphys4020}
  {\path{doi:10.1038/nphys4020}}.
\newline\urlprefix\url{www.nature.com/naturephysics}

\bibitem{Luschen2017a}
H.~P. L{\"{u}}schen, P.~Bordia, S.~S. Hodgman, M.~Schreiber, S.~Sarkar, A.~J.
  Daley, M.~H. Fischer, E.~Altman, I.~Bloch, U.~Schneider,
  \href{https://link.aps.org/doi/10.1103/PhysRevX.7.011034}{{Signatures of
  many-body localization in a controlled open quantum system}}, Physical Review
  X 7~(1) (2017) 011034.
\newblock \href {https://doi.org/10.1103/PhysRevX.7.011034}
  {\path{doi:10.1103/PhysRevX.7.011034}}.
\newline\urlprefix\url{https://link.aps.org/doi/10.1103/PhysRevX.7.011034}

\bibitem{Luschen2017c}
H.~P. L{\"{u}}schen, P.~Bordia, S.~Scherg, F.~Alet, E.~Altman, U.~Schneider,
  I.~Bloch,
  \href{https://link.aps.org/doi/10.1103/PhysRevX.7.041047}{{Observation of
  Slow Dynamics near the Many-Body Localization Transition in One-Dimensional
  Quasiperiodic Systems}}, Physical Review Letters 119~(26) (2017) 041047.
\newblock \href {https://doi.org/10.1103/PhysRevLett.119.260401}
  {\path{doi:10.1103/PhysRevLett.119.260401}}.
\newline\urlprefix\url{https://link.aps.org/doi/10.1103/PhysRevX.7.041047}

\bibitem{Xu2018}
J.~Xu, Y.~Li, \href{http://arxiv.org/abs/1811.06863
  http://dx.doi.org/10.1088/0256-307X/36/2/027201}{{Eigenstate Distribution
  Fluctuation of a Quenched Disordered Bose-Hubbard System in
  Thermal-to-Localized Transitions}}, Chinese Physics Letters 36~(2) (11 2019).
\newblock \href {https://doi.org/10.1088/0256-307X/36/2/027201}
  {\path{doi:10.1088/0256-307X/36/2/027201}}.
\newline\urlprefix\url{http://arxiv.org/abs/1811.06863
  http://dx.doi.org/10.1088/0256-307X/36/2/027201}

\bibitem{Kohlert2019}
T.~Kohlert, S.~Scherg, X.~Li, H.~P. L{\"{u}}schen, S.~Das~Sarma, I.~Bloch,
  M.~Aidelsburger,
  \href{https://0-journals-aps-org.pugwash.lib.warwick.ac.uk/prl/pdf/10.1103/PhysRevLett.122.170403}{{Observation
  of Many-Body Localization in a One-Dimensional System with a Single-Particle
  Mobility Edge}}, Physical Review Letters 122~(17) (2019).
\newblock \href {https://doi.org/10.1103/PhysRevLett.122.170403}
  {\path{doi:10.1103/PhysRevLett.122.170403}}.
\newline\urlprefix\url{https://0-journals-aps-org.pugwash.lib.warwick.ac.uk/prl/pdf/10.1103/PhysRevLett.122.170403}

\bibitem{Rispolietal2019}
M.~Rispoli, A.~Lukin, R.~Schittko, S.~Kim, M.~E. Tai, J.~L{\'e}onard,
  M.~Greiner, \href{https://doi.org/10.1038/s41586-019-1527-2}{Quantum critical
  behaviour at the many-body localization transition}, Nature 573~(7774) (2019)
  385--389.
\newblock \href {https://doi.org/10.1038/s41586-019-1527-2}
  {\path{doi:10.1038/s41586-019-1527-2}}.
\newline\urlprefix\url{https://doi.org/10.1038/s41586-019-1527-2}

\bibitem{chiaroetal2020}
B.~Chiaro, C.~Neill, A.~Bohrdt, M.~Filippone, F.~Arute, K.~Arya, R.~Babbush,
  D.~Bacon, J.~Bardin, R.~Barends, S.~Boixo, D.~Buell, B.~Burkett, Y.~Chen,
  Z.~Chen, R.~Collins, A.~Dunsworth, E.~Farhi, A.~Fowler, B.~Foxen, C.~Gidney,
  M.~Giustina, M.~Harrigan, T.~Huang, S.~Isakov, E.~Jeffrey, Z.~Jiang,
  D.~Kafri, K.~Kechedzhi, J.~Kelly, P.~Klimov, A.~Korotkov, F.~Kostritsa,
  D.~Landhuis, E.~Lucero, J.~McClean, X.~Mi, A.~Megrant, M.~Mohseni, J.~Mutus,
  M.~McEwen, O.~Naaman, M.~Neeley, M.~Niu, A.~Petukhov, C.~Quintana, N.~Rubin,
  D.~Sank, K.~Satzinger, A.~Vainsencher, T.~White, Z.~Yao, P.~Yeh, A.~Zalcman,
  V.~Smelyanskiy, H.~Neven, S.~Gopalakrishnan, D.~Abanin, M.~Knap, J.~Martinis,
  P.~Roushan, Direct measurement of non-local interactions in the many-body
  localized phase (2020).
\newblock \href {http://arxiv.org/abs/1910.06024} {\path{arXiv:1910.06024}}.

\bibitem{Guo2020}
Q.~Guo, C.~Cheng, Z.-H. Sun, Z.~Song, H.~Li, Z.~Wang, W.~Ren, H.~Dong,
  D.~Zheng, Y.-R. Zhang, R.~Mondaini, H.~Fan, H.~Wang,
  \href{https://doi.org/10.1038/s41567-020-1035-1}{Observation of
  energy-resolved many-body localization}, Nature Physics (Sep 2020).
\newblock \href {https://doi.org/10.1038/s41567-020-1035-1}
  {\path{doi:10.1038/s41567-020-1035-1}}.
\newline\urlprefix\url{https://doi.org/10.1038/s41567-020-1035-1}

\bibitem{Benini2020}
L.~Benini, P.~Naldesi, R.~A. Roemer, T.~Roscilde,
  \href{http://iopscience.iop.org/article/10.1088/1367-2630/abdf9d}{Loschmidt
  echo singularities as dynamical signatures of strongly localized phases}, New
  Journal of Physics (2021).
\newline\urlprefix\url{http://iopscience.iop.org/article/10.1088/1367-2630/abdf9d}

\bibitem{Peres1984}
A.~Peres, \href{https://link.aps.org/doi/10.1103/PhysRevA.30.1610}{{Stability
  of quantum motion in chaotic and regular systems}}, Physical Review A 30~(4)
  (1984) 1610--1615.
\newblock \href {https://doi.org/10.1103/PhysRevA.30.1610}
  {\path{doi:10.1103/PhysRevA.30.1610}}.
\newline\urlprefix\url{https://link.aps.org/doi/10.1103/PhysRevA.30.1610}

\bibitem{Jalabert2001}
R.~A. Jalabert, H.~M. Pastawski,
  \href{https://link.aps.org/doi/10.1103/PhysRevLett.86.2490}{{Environment-independent
  decoherence rate in classically chaotic systems}}, Physical Review Letters
  86~(12) (2001) 2490--2493.
\newblock \href {https://doi.org/10.1103/PhysRevLett.86.2490}
  {\path{doi:10.1103/PhysRevLett.86.2490}}.
\newline\urlprefix\url{https://link.aps.org/doi/10.1103/PhysRevLett.86.2490}

\bibitem{Iyer2013a}
S.~Iyer, V.~Oganesyan, G.~Refael, D.~A. Huse,
  \href{https://0-journals-aps-org.pugwash.lib.warwick.ac.uk/prb/pdf/10.1103/PhysRevB.87.134202}{{Many-body
  localization in a quasiperiodic system}}, Physical Review B 87~(13) (2013)
  134202.
\newblock \href {https://doi.org/10.1103/PhysRevB.87.134202}
  {\path{doi:10.1103/PhysRevB.87.134202}}.
\newline\urlprefix\url{https://0-journals-aps-org.pugwash.lib.warwick.ac.uk/prb/pdf/10.1103/PhysRevB.87.134202}

\bibitem{Naldesi2016}
P.~Naldesi, E.~Ercolessi, T.~Roscilde,
  \href{http://dx.doi.org/10.21468/SciPostPhys.1.1.010}{{Detecting a many-body
  mobility edge with quantum quenches}}, SciPost Phys 1~(1) (2016) 10.
\newblock \href {https://doi.org/10.21468/SciPostPhys.1.1.010}
  {\path{doi:10.21468/SciPostPhys.1.1.010}}.
\newline\urlprefix\url{http://dx.doi.org/10.21468/SciPostPhys.1.1.010}

\bibitem{Lee2017}
M.~Lee, T.~R. Look, S.~P. Lim, D.~N. Sheng,
  \href{https://link.aps.org/doi/10.1103/PhysRevB.96.075146}{Many-body
  localization in spin chain systems with quasiperiodic fields}, Phys. Rev. B
  96 (2017) 075146.
\newblock \href {https://doi.org/10.1103/PhysRevB.96.075146}
  {\path{doi:10.1103/PhysRevB.96.075146}}.
\newline\urlprefix\url{https://link.aps.org/doi/10.1103/PhysRevB.96.075146}

\bibitem{Nag2017}
S.~Nag, A.~Garg,
  \href{https://link.aps.org/doi/10.1103/PhysRevB.96.060203}{Many-body mobility
  edges in a one-dimensional system of interacting fermions}, Phys. Rev. B 96
  (2017) 060203.
\newblock \href {https://doi.org/10.1103/PhysRevB.96.060203}
  {\path{doi:10.1103/PhysRevB.96.060203}}.
\newline\urlprefix\url{https://link.aps.org/doi/10.1103/PhysRevB.96.060203}

\bibitem{Khemani2017}
V.~Khemani, D.~Sheng, D.~A. Huse,
  \href{https://link.aps.org/doi/10.1103/PhysRevLett.119.075702}{{Two
  Universality Classes for the Many-Body Localization Transition}}, Physical
  Review Letters 119~(7) (2017) 075702.
\newblock \href {https://doi.org/10.1103/PhysRevLett.119.075702}
  {\path{doi:10.1103/PhysRevLett.119.075702}}.
\newline\urlprefix\url{https://link.aps.org/doi/10.1103/PhysRevLett.119.075702}

\bibitem{Setiawan2017}
F.~Setiawan, D.-L. Deng, J.~H. Pixley,
  \href{https://link.aps.org/doi/10.1103/PhysRevB.96.104205}{Transport
  properties across the many-body localization transition in quasiperiodic and
  random systems}, Phys. Rev. B 96 (2017) 104205.
\newblock \href {https://doi.org/10.1103/PhysRevB.96.104205}
  {\path{doi:10.1103/PhysRevB.96.104205}}.
\newline\urlprefix\url{https://link.aps.org/doi/10.1103/PhysRevB.96.104205}

\bibitem{zhang2018}
S.-X. Zhang, H.~Yao,
  \href{https://0-link-aps-org.pugwash.lib.warwick.ac.uk/doi/10.1103/PhysRevLett.121.206601}{Universal
  properties of many-body localization transitions in quasiperiodic systems},
  Phys. Rev. Lett. 121 (2018) 206601.
\newblock \href {https://doi.org/10.1103/PhysRevLett.121.206601}
  {\path{doi:10.1103/PhysRevLett.121.206601}}.
\newline\urlprefix\url{https://0-link-aps-org.pugwash.lib.warwick.ac.uk/doi/10.1103/PhysRevLett.121.206601}

\bibitem{Znidaric2018b}
M.~Znidaric, M.~Ljubotina,
  \href{https://www.pnas.org/content/115/18/4595}{Interaction instability of
  localization in quasiperiodic systems}, Proceedings of the National Academy
  of Sciences 115~(18) (2018) 4595--4600.
\newblock \href
  {http://arxiv.org/abs/https://www.pnas.org/content/115/18/4595.full.pdf}
  {\path{arXiv:https://www.pnas.org/content/115/18/4595.full.pdf}}, \href
  {https://doi.org/10.1073/pnas.1800589115}
  {\path{doi:10.1073/pnas.1800589115}}.
\newline\urlprefix\url{https://www.pnas.org/content/115/18/4595}

\bibitem{Weiner2019}
F.~Weiner, F.~Evers, S.~Bera,
  \href{https://link.aps.org/doi/10.1103/PhysRevB.100.104204}{Slow dynamics and
  strong finite-size effects in many-body localization with random and
  quasiperiodic potentials}, Phys. Rev. B 100 (2019) 104204.
\newblock \href {https://doi.org/10.1103/PhysRevB.100.104204}
  {\path{doi:10.1103/PhysRevB.100.104204}}.
\newline\urlprefix\url{https://link.aps.org/doi/10.1103/PhysRevB.100.104204}

\bibitem{Doggen2019}
E.~V.~H. Doggen, A.~D. Mirlin,
  \href{https://link.aps.org/doi/10.1103/PhysRevB.100.104203}{Many-body
  delocalization dynamics in long aubry-andr\'e quasiperiodic chains}, Phys.
  Rev. B 100 (2019) 104203.
\newblock \href {https://doi.org/10.1103/PhysRevB.100.104203}
  {\path{doi:10.1103/PhysRevB.100.104203}}.
\newline\urlprefix\url{https://link.aps.org/doi/10.1103/PhysRevB.100.104203}

\bibitem{Shenglong2019}
S.~Xu, X.~Li, Y.-T. Hsu, B.~Swingle, S.~Das~Sarma,
  \href{https://link.aps.org/doi/10.1103/PhysRevResearch.1.032039}{Butterfly
  effect in interacting aubry-andre model: Thermalization, slow scrambling, and
  many-body localization}, Phys. Rev. Research 1 (2019) 032039.
\newblock \href {https://doi.org/10.1103/PhysRevResearch.1.032039}
  {\path{doi:10.1103/PhysRevResearch.1.032039}}.
\newline\urlprefix\url{https://link.aps.org/doi/10.1103/PhysRevResearch.1.032039}

\bibitem{Luitz2015a}
D.~J. Luitz, N.~Laflorencie, F.~Alet,
  \href{https://0-journals-aps-org.pugwash.lib.warwick.ac.uk/prb/pdf/10.1103/PhysRevB.91.081103(R)}{{Many-body
  localization edge in the random-field Heisenberg chain}}, Physical Review B
  91~(8) (2015) 81103.
\newblock \href {https://doi.org/10.1103/PhysRevB.91.081103}
  {\path{doi:10.1103/PhysRevB.91.081103}}.
\newline\urlprefix\url{https://0-journals-aps-org.pugwash.lib.warwick.ac.uk/prb/pdf/10.1103/PhysRevB.91.081103(R)}

\bibitem{Jordan1928}
P.~Jordan, E.~Wigner, {\"U}ber das paulische {\"a}quivalenzverbot, Zeitschrift
  f{\"u}r Physik 47~(9) (1928) 631--651.
\newblock \href {https://doi.org/10.1007/BF01331938}
  {\path{doi:10.1007/BF01331938}}.

\bibitem{Weinberg2017QuSpin:Chains}
P.~Weinberg, M.~Bukov, {QuSpin: a Python package for dynamics and exact
  diagonalisation of quantum many body systems part I: spin chains}, SciPost
  Physics 2~(1) (2017) 3.
\newblock \href {https://doi.org/10.21468/scipostphys.2.1.003}
  {\path{doi:10.21468/scipostphys.2.1.003}}.

\bibitem{Weinberg2018}
P.~Weinberg, M.~Bukov,
  \href{https://scipost.org/10.21468/SciPostPhys.7.2.020}{{QuSpin: a Python
  Package for Dynamics and Exact Diagonalisation of Quantum Many Body Systems.
  Part II: bosons, fermions and higher spins}}, SciPost Phys. 7 (2019) 20.
\newblock \href {https://doi.org/10.21468/SciPostPhys.7.2.020}
  {\path{doi:10.21468/SciPostPhys.7.2.020}}.
\newline\urlprefix\url{https://scipost.org/10.21468/SciPostPhys.7.2.020}

\bibitem{Aubry1980}
S.~Aubry, G.~Andr{\'{e}}, {Analyticity breaking and Anderson localization in
  incommensurate lattices} (1980).

\bibitem{Heyl2013}
M.~Heyl, A.~Polkovnikov, S.~Kehrein,
  \href{https://0-journals-aps-org.pugwash.lib.warwick.ac.uk/prl/pdf/10.1103/PhysRevLett.110.135704
  https://link.aps.org/doi/10.1103/PhysRevLett.110.135704}{{Dynamical Quantum
  Phase Transitions in the Transverse-Field Ising Model}}, Physical Review
  Letters 110~(13) (2013) 135704.
\newblock \href {https://doi.org/10.1103/PhysRevLett.110.135704}
  {\path{doi:10.1103/PhysRevLett.110.135704}}.
\newline\urlprefix\url{https://0-journals-aps-org.pugwash.lib.warwick.ac.uk/prl/pdf/10.1103/PhysRevLett.110.135704
  https://link.aps.org/doi/10.1103/PhysRevLett.110.135704}

\bibitem{Heyl2014}
M.~Heyl, \href{http://arxiv.org/abs/1403.4570
  http://dx.doi.org/10.1103/PhysRevLett.113.205701
  https://link.aps.org/doi/10.1103/PhysRevLett.113.205701}{{Dynamical quantum
  phase transitions in systems with broken-symmetry phases}}, Physical Review
  Letters 113~(20) (2014) 205701.
\newblock \href {https://doi.org/10.1103/PhysRevLett.113.205701}
  {\path{doi:10.1103/PhysRevLett.113.205701}}.
\newline\urlprefix\url{http://arxiv.org/abs/1403.4570
  http://dx.doi.org/10.1103/PhysRevLett.113.205701
  https://link.aps.org/doi/10.1103/PhysRevLett.113.205701}

\bibitem{Jurcevic2017}
P.~Jurcevic, H.~Shen, P.~Hauke, C.~Maier, T.~Brydges, C.~Hempel, B.~P. Lanyon,
  M.~Heyl, R.~Blatt, C.~F. Roos,
  \href{https://0-journals-aps-org.pugwash.lib.warwick.ac.uk/prl/pdf/10.1103/PhysRevLett.119.080501
  https://link.aps.org/doi/10.1103/PhysRevLett.119.080501}{{Direct Observation
  of Dynamical Quantum Phase Transitions in an Interacting Many-Body System}},
  Physical Review Letters 119~(8) (2017) 080501.
\newblock \href {https://doi.org/10.1103/PhysRevLett.119.080501}
  {\path{doi:10.1103/PhysRevLett.119.080501}}.
\newline\urlprefix\url{https://0-journals-aps-org.pugwash.lib.warwick.ac.uk/prl/pdf/10.1103/PhysRevLett.119.080501
  https://link.aps.org/doi/10.1103/PhysRevLett.119.080501}

\bibitem{Yang2017}
C.~Yang, Y.~Wang, P.~Wang, X.~Gao, S.~Chen, {Dynamical signature of
  localization-delocalization transition in a one-dimensional incommensurate
  lattice}, Physical Review B 95~(18) (5 2017).
\newblock \href {https://doi.org/10.1103/PhysRevB.95.184201}
  {\path{doi:10.1103/PhysRevB.95.184201}}.

\bibitem{Heyl2018}
M.~Heyl, \href{https://doi.org/10.1088%2F1361-6633%2Faaaf9a}{Dynamical quantum
  phase transitions: a review}, Reports on Progress in Physics 81~(5) (2018)
  054001.
\newblock \href {https://doi.org/10.1088/1361-6633/aaaf9a}
  {\path{doi:10.1088/1361-6633/aaaf9a}}.
\newline\urlprefix\url{https://doi.org/10.1088%2F1361-6633%2Faaaf9a}

\bibitem{Guoetal2019}
X.-Y. Guo, C.~Yang, Y.~Zeng, Y.~Peng, H.-K. Li, H.~Deng, Y.-R. Jin, S.~Chen,
  D.~Zheng, H.~Fan,
  \href{https://link.aps.org/doi/10.1103/PhysRevApplied.11.044080}{Observation
  of a dynamical quantum phase transition by a superconducting qubit
  simulation}, Phys. Rev. Applied 11 (2019) 044080.
\newblock \href {https://doi.org/10.1103/PhysRevApplied.11.044080}
  {\path{doi:10.1103/PhysRevApplied.11.044080}}.
\newline\urlprefix\url{https://link.aps.org/doi/10.1103/PhysRevApplied.11.044080}

\bibitem{Yin2018}
H.~Yin, S.~Chen, X.~Gao, P.~Wang,
  \href{https://link.aps.org/doi/10.1103/PhysRevA.97.033624}{{Zeros of
  Loschmidt echo in the presence of Anderson localization}}, Physical Review A
  97~(3) (2018) 033624.
\newblock \href {https://doi.org/10.1103/PhysRevA.97.033624}
  {\path{doi:10.1103/PhysRevA.97.033624}}.
\newline\urlprefix\url{https://link.aps.org/doi/10.1103/PhysRevA.97.033624}

\bibitem{Halimeh2019}
J.~C. Halimeh, N.~Yegovtsev, V.~Gurarie,
  \href{http://arxiv.org/abs/1903.03109}{{Dynamical quantum phase transitions
  in many-body localized systems}}, arxiv:1903.03109 (2019).
\newline\urlprefix\url{http://arxiv.org/abs/1903.03109}

\bibitem{Schuster2002}
C.~Schuster, R.~A. R\"omer, M.~Schreiber,
  \href{https://link.aps.org/doi/10.1103/PhysRevB.65.115114}{Interacting
  particles at a metal-insulator transition}, Phys. Rev. B 65 (2002) 115114.
\newblock \href {https://doi.org/10.1103/PhysRevB.65.115114}
  {\path{doi:10.1103/PhysRevB.65.115114}}.
\newline\urlprefix\url{https://link.aps.org/doi/10.1103/PhysRevB.65.115114}

\bibitem{Serbyn2013UniversalSystems}
M.~Serbyn, Z.~Papi{\'{c}}, D.~A. Abanin, {Universal slow growth of entanglement
  in interacting strongly disordered systems}, Physical Review Letters 110~(26)
  (2013) 1--5.
\newblock \href {https://doi.org/10.1103/PhysRevLett.110.260601}
  {\path{doi:10.1103/PhysRevLett.110.260601}}.

\bibitem{Bera2015a}
S.~Bera, H.~Schomerus, F.~Heidrich-Meisner, J.~H. Bardarson,
  \href{https://0-journals-aps-org.pugwash.lib.warwick.ac.uk/prl/pdf/10.1103/PhysRevLett.115.046603}{{Many-Body
  Localization Characterized from a One-Particle Perspective}}, Physical Review
  Letters 115~(4) (2015).
\newblock \href {https://doi.org/10.1103/PhysRevLett.115.046603}
  {\path{doi:10.1103/PhysRevLett.115.046603}}.
\newline\urlprefix\url{https://0-journals-aps-org.pugwash.lib.warwick.ac.uk/prl/pdf/10.1103/PhysRevLett.115.046603}

\bibitem{ScullyBook1997}
M.~O. Scully, M.~S. Zubairy, Quantum Optics, Cambridge University Press, 1997,
  Ch.~6.
\newblock \href {https://doi.org/10.1017/CBO9780511813993}
  {\path{doi:10.1017/CBO9780511813993}}.

\bibitem{Sierant2017a}
P.~Sierant, D.~Delande, J.~Zakrzewski,
  \href{https://link.aps.org/doi/10.1103/PhysRevA.95.021601}{Many-body
  localization due to random interactions}, Phys. Rev. A 95 (2017) 021601.
\newblock \href {https://doi.org/10.1103/PhysRevA.95.021601}
  {\path{doi:10.1103/PhysRevA.95.021601}}.
\newline\urlprefix\url{https://link.aps.org/doi/10.1103/PhysRevA.95.021601}

\bibitem{Sierant2017b}
P.~Sierant, D.~Delande, J.~Zakrzewski, Many-body localization for randomly
  interacting bosons, Acta Physica Polonica A 132 (07 2017).
\newblock \href {https://doi.org/10.12693/APhysPolA.132.1707}
  {\path{doi:10.12693/APhysPolA.132.1707}}.

\bibitem{Janarek2018}
J.~Janarek, D.~Delande, J.~Zakrzewski,
  \href{https://link.aps.org/doi/10.1103/PhysRevB.97.155133}{Discrete disorder
  models for many-body localization}, Phys. Rev. B 97 (2018) 155133.
\newblock \href {https://doi.org/10.1103/PhysRevB.97.155133}
  {\path{doi:10.1103/PhysRevB.97.155133}}.
\newline\urlprefix\url{https://link.aps.org/doi/10.1103/PhysRevB.97.155133}

\bibitem{Guarrera2007}
V.~Guarrera, L.~Fallani, J.~E. Lye, C.~Fort, M.~Inguscio,
  \href{http://stacks.iop.org/1367-2630/9/i=4/a=107?key=crossref.498f1ead89a32b29c85097f09710baa8}{{Inhomogeneous
  broadening of a Mott insulator spectrum}}, New Journal of Physics 9~(4)
  (2007) 107--107.
\newblock \href {https://doi.org/10.1088/1367-2630/9/4/107}
  {\path{doi:10.1088/1367-2630/9/4/107}}.
\newline\urlprefix\url{http://stacks.iop.org/1367-2630/9/i=4/a=107?key=crossref.498f1ead89a32b29c85097f09710baa8}

\bibitem{Belitz1994}
D.~Belitz, T.~R. Kirkpatrick,
  \href{https://link.aps.org/doi/10.1103/RevModPhys.66.261}{{The Anderson-Mott
  transition}}, Rev. Mod. Phys. 66~(2) (1994) 261--380.
\newblock \href {https://doi.org/10.1103/RevModPhys.66.261}
  {\path{doi:10.1103/RevModPhys.66.261}}.
\newline\urlprefix\url{https://link.aps.org/doi/10.1103/RevModPhys.66.261}

\end{thebibliography}

\end{document}